\documentclass[
reprint,
nofootinbib,
amsmath,amssymb,amsbsy,
prd,
twocolumn,
superscriptaddress,
longbibliography,
preprintnumbers]{revtex4-2}

\usepackage{relsize}

\usepackage[acronyms, nohypertypes={acronym}, nopostdot, style=super, nonumberlist, toc]{glossaries}
\glsenableentrycount
\setacronymstyle{long-sc-short}

\newacronym{WF}{wf}{Wilson-Fisher}
\newacronym{AF}{af}{asymptotically free}
\newacronym{RG}{rg}{renormalization group}
\newacronym{WZW}{wzw}{Wess-Zumino-Witten}
\newacronym[longplural={conformal field theories}]{CFT}{cft}{conformal field theory}
\newacronym[longplural={lattice field theories}]{LFT}{lft}{lattice field theory}
\newacronym[longplural={effective field theories}]{EFT}{eft}{effective field theory}
\newacronym[longplural={quantum field theories}]{QFT}{qft}{quantum field theory}
\newacronym{LEC}{lec}{low-energy constant}
\newacronym{QCD}{qcd}{quantum chromodynamics}
\newacronym{MC}{mc}{Monte Carlo}
\newacronym{IR}{ir}{infrared}
\newacronym{UV}{uv}{ultraviolet}
\newacronym{SNR}{snr}{signal-to-noise ratio}
\newacronym{NLSM}{nl$\sigma$m}{nonlinear sigma model}
\newacronym{PCM}{pcm}{principal chiral model}
\newacronym{CSA}{csa}{Cartan subalgebra}
\newacronym{SSB}{ssb}{spontaneous symmetry breaking}
\newacronym{DOF}{dof}{degrees of freedom}
\newacronym{DMRG}{dmrg}{densiy matrix renormalization group}
\newacronym{YM}{ym}{Yang-Mills}
\newacronym{QLM}{qlm}{quantum link model}
\newacronym{KG}{kg}{Kogut-Susskind}
\newacronym{KG-QLM}{kg-qlm}{Kogut-Susskind quantum link model}
\newacronym{D-QLM}{d-qlm}{D-theory quantum link model}
\newacronym{SPT}{spt}{symmetry protected topological} 
\newacronym{GW}{gw}{Ginsparg-Wilson}
\newacronym{FK}{fk}{Fidkowski-Kitaev}
\newacronym{CS}{cs}{Chern-Simons}
\newacronym{APS}{aps}{Atiyah-Patodi-Singer}
\newacronym{PV}{pv}{Pauli-Villars}

\usepackage{mathtools}
\usepackage{amsmath,amssymb,amsbsy,amsfonts}

\usepackage{hyperref}
\hypersetup{
    unicode,
    pdfprintscaling=None,
    colorlinks,
    breaklinks
}
\hypersetup{
  colorlinks=true,   
  linkcolor=cyan,    
  citecolor=blue,    
  filecolor=magenta, 
  urlcolor=blue      
}

\usepackage[capitalise]{cleveref}
\crefname{section}{Sec.}{Secs.}

\usepackage{microtype}

\newcommand\del\partial

\usepackage{dutchcal}

\newcommand{\ole}{\overline}

\usepackage{makecell}

\newcommand{\conD}{D}
\newcommand{\gwD}{\mathcal{D}}

\usepackage{enumitem}
\usepackage{pifont}

\usepackage{graphicx}
\usepackage{amsfonts}
\usepackage[named]{xcolor}
\usepackage{tocvsec2}
\usepackage{slashed}
\usepackage{cancel}
\usepackage{comment}

\newcommand\beq{\begin{eqnarray}}
\newcommand\eeq{\end{eqnarray}}

\newcommand{\mybar}[1]{\kern 0.6pt\overline{\kern -0.6pt#1\kern -0.6pt}\kern 0.6pt}

\def\U\Omega{U(1)_{\Omega}}

\usepackage[section]{placeins}

\begin{document}

\title{Toward A Ginsparg-Wilson Lattice Hamiltonian}

\author{Michael Clancy}
\email{mclancy2@uw.edu}
\affiliation{Institute for Nuclear Theory, Box 351550, Seattle WA 98195-1550}
\preprint{INT-PUB-23-051}

\begin{abstract}
To address quantum computation of quantities in quantum chromodynamics (QCD) for which chiral symmetry is important, it would be useful to have the Hamiltonian for a fermion satisfying the Ginsparg-Wilson (GW) equation. I work with an approximate solution to the GW equation which is fractional linear in time derivatives. The resulting Hamiltonian is non-local and has ghosts, but is free of doublers and has the correct continuum limit. This construction works in general odd spatial dimensions, and I provide an explicit expression for the Hamiltonian in 1 spatial dimension.  
\end{abstract}

\maketitle

\section{Introduction}
\label{sec:intro}
There are a number of computations in QCD that require a good realization of chiral symmetry. These include the color-flavor-locking phase \cite{Alford_1999} and the chiral symmetry restoring phase transition \cite{pisarski2003review}, both of which one would hope to be able to study on the lattice. The overlap operator \cite{Narayanan:1993ss,Narayanan:1993sk} in the Euclidean Lagrangian formulation offers the ideal realization of lattice chiral symmetry in the form of L\"{u}scher symmetry \cite{Luscher:1998pqa}, a lattice symmetry which tends toward chiral symmetry in the continuum limit. However, computations using the path integral are afflicted by sign problems. A Hamiltonian approach on a quantum computer might be able to solve these issues, but there currently does not exist a Hamiltonian for GW fermions.

Chiral symmetry can be expected to fail on the lattice because the lattice spacing introduces a mass scale, and masses violate chiral symmetry. This can be made more precise by the Nielsen-Ninomiya no-go theorem \cite{Nielsen:1980rz}; there is no lattice Dirac operator $\gwD$ in 4 spacetime dimensions which has chiral symmetry, i.e. satisfies
\begin{equation} \label{chiralsym}
\{\gamma_5,\gwD\} = 0,
\end{equation}
and has other desirable features, namely the correct continuum limit, freedom from doublers, and locality \cite{Luscher:1998pqa}. Ginsparg and Wilson \cite{Ginsparg:1981bj} suggested that this should be replaced by:
\begin{equation} \label{GWOG}
\{\gamma_5,\gwD\} = a \gwD \gamma_5 \gwD,
\end{equation}
so that exact chiral symmetry fails at the order of the lattice spacing $a$. The first method for putting chiral fermions on the lattice involved edge states of a domain wall defect in one higher dimension \cite{Kaplan:1992bt}. Neuberger and Narayanan \cite{Narayanan:1993ss,Narayanan:1993sk} found that this system could be studied in 4 dimensions via the ``overlap'' operator,
\begin{equation} \label{1PV}
\gwD  = \frac{M}{2} \left( 1 + V \right), \quad V  =  \frac{D_w}{\sqrt{D_w^{\dagger} D_w}},
\end{equation}
where $M = 1/a$ is the inverse lattice spacing, and $D_w$ is the 4-dimensional Wilson Dirac operator, which, in the absence of gauge fields, can be written in momentum space as:
\begin{equation} \label{DW4}
D_w = i \sum_{\mu =1}^4 \gamma^\mu \sin( p_\mu /M) - 1 + \sum_{\mu=1}^4 (1 - \cos (p_\mu/M)).
\end{equation}
This operator has the correct continuum limit, and is not hermitian, but is instead ``$\gamma_5$-hermitian'':
\begin{equation} \label{g5herm}
\gamma_5 \gwD \gamma_5 = \gwD^{\dagger}.
\end{equation}
In fact, it can be quickly checked that any operator of the form
\begin{equation} \label{GWGEN}
\gwD = \frac{M}{2} \left( 
1 + V
\right); 
\quad \gamma_5 V \gamma_5 = V^\dagger, \quad V^\dagger V = I,
\end{equation} 
satisfies Eq. (\ref{GWOG}) \cite{Neuberger:1997fp}. In this case I say $\gwD$ is an overlap operator, though in general it may not necessarily be constructed in terms of a state overlap. L\"{u}scher \cite{Luscher:1998pqa} first observed that this operator has the following symmetry:
\begin{equation} \label{lusc}
\delta \psi = \gamma_5 \left( \frac{1-V}{2} \right) \psi, \quad \delta \ole{\psi} = \ole{\psi} \left( \frac{1-V}{2} \right) \gamma_5.
\end{equation}
In the continuum limit, this becomes chiral symmetry. L\"{u}scher noted that the Jacobian of this transformation produces the index of $\gwD$, a lattice version of the Fujikawa calculation \cite{fujikawa1979path,Luscher:1998pqa} of the chiral anomaly. Indeed, there is a good deal of freedom in defining this L\"{u}scher symmetry; hereafter I will refer to any symmetry
\begin{equation}
\delta \psi = \Gamma \psi, \quad \delta \ole{\psi} = \ole{\psi} \ole{\Gamma},
\end{equation}
for which
\begin{equation}
\lim_{M \to \infty} \Gamma = \lim_{M \to \infty} \ole{\Gamma} = \gamma_5,
\end{equation}
and whose determinant reproduces the index of $\gwD$, as a L\"{u}scher symmetry \cite{Luscher:1998pqa}.

To find a Hamiltonian describing a GW fermion, one may try to compute the transfer matrix of Eq. (\ref{1PV}) directly, but this involves square roots of the time derivative and is therefore challenging. Creutz et al. \cite{Creutz:2001wp} considered the following construction. First, define the 3-dimensional overlap operator 
\begin{equation} \label{d3}
d = \frac{M}{2} \left( 1 + v \right), \quad v = \frac{d_w}{\sqrt{d_w^\dagger d_w}},
\end{equation}
where  $d_w$ is the 3-dimensional analogue of Eq. (\ref{DW4}):
\begin{equation} \label{DW3}
d_w = i \sum_{i =1}^3 \gamma^i \sin ( p_i/M) - M + \sum_{i=1}^3 (1 - \cos (p_i/M)),
\end{equation}
and $\gamma^i$ are $4 \times 4$ Clifford algebra matrices. Then by analogy with the continuum Hamiltonian
\begin{equation}
H_\psi^{c} = \int d^3 x \psi^\dagger i \gamma^0 \gamma^i D_i \psi,
\end{equation}
(where $D_i,H_\psi^{c}$ denote the continuum covariant derivative and continuum Hamiltonian, respectively), it is reasonable to identify $\gamma^i \conD_i$ with the 3-dimensional Dirac operator, and formulate a lattice prescription for a Hamiltonian via the replacement $\gamma^i D_i \to d$:
\begin{equation}\label{chnham}
H_\psi \equiv \psi^\dagger i \gamma^0 d \psi.
\end{equation}
This system has the symmetry of Eq. (\ref{lusc}), and associated to that symmetry is the charge
\begin{equation}
Q_5 = \psi^\dagger \gamma_5 \left(\frac{1-V}{2} \right) \psi.
\end{equation}
This chiral charge is conserved with respect to $H_\psi$, i.e. $[H_\psi,Q_5] = 0$, but upon introduction of the gauge field Hamiltonian
\begin{equation} \label{Hg}
H_{g} = \frac{1}{2} ( E^2 + B^2 ),
\end{equation}
one finds $[H_g,Q_5] \neq 0$, since $E^2$ involves derivatives with respect to the gauge fields in the quantized theory, and the $V$ appearing in $Q_5$ involves link variables.

It is important to note that the Hamiltonian considered by Creutz et al. is not derived from a GW fermion in the Euclidean Lagrangian; it is simply an ansatz. If it were, it would enjoy a full L\"{u}scher symmetry that descends to the Hamiltonian formulation, even in the presence of gauge fields. 

Therefore it is sensible to start at the level of the Lagrangian, with a modified overlap operator which still solves the GW equation, but from which the extraction of a Hamiltonian is considerably easier. It is simpler to consider a theory which is fractional linear in time derivatives, i.e. a rational expression linear in time derivatives. The feasibility of such an approach will become clear by construction of an overlap operator in the continuum with ghosts, namely a Pauli-Villars regulated fermion.

In Section \ref{sec:PVSEC}, I will describe the way in which Pauli-Villars fermions satisfy the GW relation, and the Hamiltonian and L\"{u}scher symmetry associated to them. In Section \ref{sec:LATPV:LAG} I will derive a Lagrangian describing a GW fermion in discrete space and continuous time, and generalizing the arguments of Section \ref{sec:PVSEC} I will derive a Hamiltonian describing the system. In Section \ref{sec:LATPV:HAM} I will describe the properties of this Hamiltonian. 

\section{Pauli-Villars as Overlap} \label{sec:PVSEC}
In a recent paper generalizing the GW relation \cite{CKS2023}, it was found that the GW equation holds for a Pauli-Villars regulated fermion in the continuum. I will derive the Hamiltonian for this example, as it is instructive for generalization to the lattice.

A Pauli-Villars regulated fermion is equivalent to a Lagrangian with the following Dirac operator:
\begin{equation} \label{PVOP}
\mathcal{L} = \ole{\psi} \gwD \psi, \quad \gwD = M \frac{\slashed{D}}{\slashed{D} + M},
\end{equation}
where $\slashed{D}$ is the usual Euclidean Dirac operator, $\slashed{D} = \gamma^\mu D_\mu$. This may be rewritten
\begin{equation} \label{PVform}
\gwD = \frac{M}{2} \left( 1 + V \right), \quad V = \frac{\slashed{D}/M - 1}{\slashed{D}/M + 1}.
\end{equation}
This $\gwD$ satisfies Eq. (\ref{GWGEN}), so it is an overlap operator. For reasons that will become clear shortly, is helpful to define $A = \slashed{\conD}/M - 1$, and note $V$ is of the form:
\begin{equation} \label{sauce}
V = - A^{-1} A^\dagger; \quad \gamma_5 A \gamma_5 = A^\dagger.
\end{equation}
In order to make this theory look familiar, I introduce ghost fields $\phi, \ole{\phi}$ with opposite statistics to the action, so that the full Lagrangian is
\begin{equation}
\mathcal{L}_{tot} = \ole{\psi} \frac{M}{2} A^{-1} (A-A^\dagger) \psi + \ole{\phi} \phi.
\end{equation}
I perform the simultaneous change of variables
\begin{equation} \label{PVCOV}
\ole{\psi}' = \ole{\psi}A^{-1}, \quad \ole{\phi}' = \ole{\phi} A^{-1}.
\end{equation}
This change of variables has trivial Jacobian in the path integral because of the opposite statistics. Under this change of variables the Lagrangian becomes
\begin{equation}
\mathcal{L}_{tot} = \ole{\psi}' \slashed{D} \psi + \ole{\phi}'(\slashed{D} + M) \phi.
\end{equation}
Consider how a L\"{u}scher symmetry $\Gamma,\ole{\Gamma}$ on $\psi,\ole{\psi}$ is affected by this change of variables. $\Gamma$ is unaffected, while the new $\ole{\Gamma}$ is related to the original by
\begin{equation} \label{lbarprime}
\ole{\Gamma}' = A \ole{\Gamma} A^{-1}.
\end{equation}
In particular, consider the choice:
\begin{equation} \label{lusc2}
\Gamma = \gamma_5, \quad \ole{\Gamma} = - V \gamma_5.
\end{equation}
Since $V$ is of the form Eq. (\ref{sauce}), Eq. (\ref{lbarprime}) becomes
\begin{equation}
\ole{\Gamma}' =  A^{-1} A^\dagger \gamma_5 A^{-1} =  A^\dagger \gamma_5 A^{-1} = \gamma_5 A A^{-1} = \gamma_5.
\end{equation}
In summary, the Pauli-Villars fermion described in Eq. (\ref{PVOP}), with the L\"{u}scher symmetry of Eq. (\ref{lusc2}) descends to a massless fermion with ordinary chiral symmetry and a heavy ghost fermion where the symmetry acts trivially. The Hamiltonian of the theory is thus 
\begin{equation}
H^c = H_{\psi}^c + H_{\phi}^c, 
\end{equation}
where
\begin{align}
H_\psi^c &  = \int d^3 x \psi^\dagger i \gamma^0 \gamma^i D_i \psi, \label{Hf} \\
H_\phi^g &  = \int d^3 x \phi^\dagger i \gamma^0 \gamma^i D_i \phi  - \phi^\dagger \gamma^0 M \phi.
\end{align}
In order to study the dynamics of $H_\psi^c$ alone, one must work in the vacuum to vacuum sector of the ghost theory. In the $M \to \infty$ limit, any excitations of $H_\phi^g$ are of order $M$, and so can be ignored.

\section{Combined Overlap} \label{sec:LATPV}
\subsection{Overlap Lagrangian} \label{sec:LATPV:LAG}
Now I work in continuous time and latticized space. Since the Pauli-Villars and overlap solutions apply to continuum and lattice cases of an overlap operator respectively, it is reasonable to try to write an ansatz for $\gwD$ which combines the forms of Eqs. (\ref{PVform}) and (\ref{1PV}). Such an operator is determined by a choice of unitary and $\gamma_5$-hermitian $V$. Recall the 3-dimensional analogue $v$ in Eq. (\ref{d3}). Since the low energy spectrum of $v$ is $-1 + i\slashed{\vec{p}}/M$ in the free theory, a reasonable ansatz incorporating Pauli-Villars regularization might be:
\begin{equation} \label{try1}
V = \frac{\gamma^0 \partial_t + Mv}{\gamma^0 \partial_t - Mv^\dagger}.
\end{equation}
Here when I write the quotient, I mean left-multiplication by the inverse of the denominator, as in Eq. (\ref{sauce}). Note that the $V$ of Eq. (\ref{try1}) also satisfies the relations of Eq. (\ref{sauce}). Furthermore $V$ is $\gamma_5$-hermitian, unitary, and has the correct low energy spectrum. However, this $V$ has doublers: note that zero-modes of $\gwD$ correspond to $-1$-modes of $V$, and therefore generally an overlap operator has doublers if there are any $V = -1$ modes away from the origin in the Brillouin zone. Note that at $v = 1$, $V = -1$, and so the free theory already has doublers at $\vec{p}_i = \pi/a$. This is because  at $\partial_t = 0$, $V = -v/v^\dagger$. Since complex conjugation treats the $v = \pm 1$-modes identically, doublers arise. Therefore, in order to find a $V$ without doublers, the $v = \pm 1$ modes need to be treated differently under conjugation. One way to do this is to replace $v \to -\sqrt{-v}$; I will address the complications of defining the square root shortly. Furthermore, I introduce an extra mass to ensure that the denominator stays invertible. Then $V$ becomes instead
\begin{equation} \label{tryVf}
V = \frac{\frac{1}{2} \gamma^0 \partial_t - M \sqrt{-v}-M}{\frac{1}{2} \gamma^0 \partial_t + M\sqrt{-v}^\dagger+M}.
\end{equation}
This $V$ is unitary,\footnote{In general, $V = -A^{-1} A^\dagger$ is unitary if $A$ is normal ($A A^\dagger = A^\dagger A$), which holds for all cases considered in this paper.} has the correct continuum limit, is free of doublers, and, except for the $v = 1$ modes, is $\gamma_5$-hermitian. Note that since $[V,P_1] = 0$ and $[\gamma_5,P_1] = 0$, I can restrict the L\"{u}scher symmetry of Eq. (\ref{lusc2}) to the $v \neq 1$ subspace.

I define the square root $\sqrt{U}$ of a unitary matrix $U$ generally as the unique matrix whose log spectrum lies in the interval $(-i\pi/2,i\pi/2]$, and which squares to $U$; this can be equivalently defined as the matrix whose eigenvalues are the square root of the eigenvalues of $U$, with the same eigenvectors. Such a definition involves choosing a branch cut, namely $\sqrt{-1} = i$, and therefore a discontinuity at the edge of the BZ (which introduces non-locality).  Note $\sqrt{U^\dagger} \neq \sqrt{U}^\dagger$, but instead
\begin{equation}
\sqrt{U^\dagger} = \sqrt{U}^\dagger - 2iP_{-1},
\end{equation}
where $P_{-1}$ is the projector onto the $-1$ modes of $U$. In the same vein, for $\gamma_5$-hermitian $v$, 
\begin{equation}
\gamma_0 \sqrt{-v} \gamma^0 = \gamma_5 \sqrt{-v} \gamma_5 = \sqrt{-v^\dagger}.
\end{equation}
This $V$ is therefore an approximate solution to the GW equation, with failure only at the high momentum modes $v = 1$. Since this $V$ can be written $V = -A^{-1} A^\dagger$, with $\gamma_5 A \gamma_5 = A^\dagger$, I repeat the analysis of Section\ref{sec:PVSEC}. Since $[V,P_1] = 0$ and $[\gamma_5,P_1] = 0$, any L\"{u}scher symmetries can be restricted to the $v \neq 1$ subspace. Then the Hamiltonian becomes
\begin{align} \nonumber
H & = H_\psi + H_\phi, \\ \vspace{5mm}
H_\psi & =  \psi^\dagger h_\psi \psi, \quad H_\phi =  \phi^\dagger h_\phi \phi.
\end{align}
where 
\begin{align} \nonumber
h_\psi &= M \gamma^0 \left( \sqrt{-v}^\dagger - \sqrt{-v} \right), \\
h_\phi & = 2 M \gamma^0 \sqrt{-v}^\dagger + 2M \gamma^0,
\end{align}
and repeated (suppressed) indices are summed over. The ghost fields here have been rescaled to be canonically normalized. The ghost Hamiltonian $H_\phi$ is clearly gapped, and energy levels occur at the order of the inverse lattice spacing $M$. Therefore, I consider only ghost vacuum-to-vacuum amplitudes of the combined system, and work only with $H_\psi$.

\subsection{Overlap Hamiltonian} \label{sec:LATPV:HAM}
Including the modes at $v = 1$, $h_\psi$ is not hermitian, but rather can be written as the sum of a hermitian and non-hermitian piece:
\begin{align}
h_\psi = M \gamma^0 \left( \sqrt{-v^\dagger} - \sqrt{-v} \right) - 2iM \gamma^0  P_1 .
\end{align}
In fact, chiral symmetry also fails at exactly these one-modes:
\begin{align}
\gamma_5 h_\psi \gamma_5 = h_\psi^\dagger,
\end{align}
so that $h_\psi$ is still $\gamma_5$-hermitian. Note that since $[\gamma^0,P_1] = 0$, these non-hermitian modes have both positive and negative imaginary eigenvalues, corresponding to states that blow up and decay, respectively. However, $[h_\psi,P_1] = 0$, so the space of states may be restricted to the subspace $v \neq 1$. In this subspace, $h_\psi$ is hermitian, and has chiral symmetry. Furthermore, it is easily checked that it has the right continuum limit, since $v \to -1 + i \slashed{p}$. The same holds true in the presence of gauge fields which are sufficiently smooth. 

It is illuminating to consider replicating this construction in $d = 1 + 1$. One finds $v = - e^{-ip\gamma_1}$, and in this case the free Hamiltonian is explicitly
\begin{equation}
h_\psi = 2M\gamma_\chi \sin{p/2M}, 
\end{equation}
where $\gamma_\chi = i \gamma_1 \gamma_2$ is the 3-dimensional analogue of $\gamma_5$. This matches the continuum Hamiltonian for a free 2-component Dirac spinor in the low energy limit.

In the quantized theory, it is obvious that the chiral charge is conserved with respect to $h_\psi$. The chiral charge may be written
\begin{equation} \label{Q5lat}
Q_5 = \psi^\dagger \gamma_5 \psi.
\end{equation}
In the quantized theory, $[H_\psi,Q_5] = 0$ since $[h_\psi,\gamma_5] = 0$. The gauge field Hamiltonian $h_g = E^2 + B^2$ trivially commutes with $\gamma_5$, so that the chiral charge is conserved in the full theory. This is in contrast to the Creutz et al. \cite{Creutz:2001wp} Hamiltonian of Eq. (\ref{chnham}), since the chiral charge in Eq. (\ref{Q5lat}) no longer involves the overlap operator, but is in direct analogy with the continuum chiral charge. 
\section{Conclusions}
I have derived a Hamiltonian for a massless spatial-lattice fermion with exact chiral symmetry from a spatial-lattice continuous-time Lagrangian for an approximate GW fermion, starting with an overlap operator with L\"{u}scher symmetry and introducing ghosts. This came at the expense of locality, and unitarity at $v = 1$. However, the physical complications of this sickness are not clear. In the usual overlap operator in Eq. (\ref{1PV}), the $V = 1$ modes have physical significance, as they must compensate for the index of the zero-modes \cite{Luscher:1998pqa}. However, for the overlap operator determined by Eq. (\ref{tryVf}), the $v = 1$ modes do not correspond to $V = 1$ modes, and so they appear to be modes at the cutoff which do not affect the low energy physics.

It is worth noting that the non-hermiticity of the $v = 1$ modes can be eliminated by replacing $\sqrt{-v}^\dagger \to \sqrt{-v^\dagger}$ in Eq. (\ref{tryVf}), but in the Hamiltonian language this introduces extra zero-modes at $v = 1$, i.e. doublers.

One of the main motivations for this letter is demonstrating the difficulty of the formulation of a consistent Hamiltonian describing GW fermions. It is my hope that this either spurs interest in a solution that does not suffer the shortcomings of the Hamiltonian considered in this letter, or in a no-go theorem that forbids the formulation of such a Hamiltonian. 

\section{Acknowledgements}
I thank David B. Kaplan and Hersh Singh for useful discussions. This research is supported in part by DOE Grant No. DE-FG02-00ER41132.
\bibliography{refs}
\end{document}